\documentclass[micromachines,article,accept,moreauthors,pdftex]{Definitions/mdpi} 

\usepackage{ upgreek }
 \usepackage[labelformat=simple]{subcaption}

\DeclareCaptionLabelFormat{subcaptionlabel}{\normalfont(\textbf{#2}\normalfont)}
\firstpage{1} 
\pubvolume{13}
\issuenum{13}
\articlenumber{1145}
\pubyear{2022}
\copyrightyear{2022}
\externaleditor{{Academic Editor:  Saulius Juodkazis}} 

\datereceived{15 June 2022} 
\dateaccepted{17 June 2022} 
\datepublished{19 June 2022} 
\hreflink{https://doi.org/10.3390/mi13071145}

\usepackage[italian,english]{babel}
\usepackage[utf8]{inputenc}
\usepackage{latexsym,graphicx,xcolor,tabularx,colortbl,tikz,soul}
\usepackage[separate-uncertainty=true]{siunitx}
\DeclareSIUnit\sq{\ensuremath{\Box}}
\DeclareSIUnit\cell{cell}
\usepackage[version=4]{mhchem}

\usepackage{caption}
\usepackage{subcaption}

\graphicspath{{./images/}}

\newcommand{\secref}[1]{Section~\ref{#1}}
\newcommand{\figref}[1]{\figurename~\ref{#1}}
\newcommand{\tabref}[1]{\tablename~\ref{#1}}

\Title{\textls[-25]{Toward Higher Integration Density in Femtosecond-Laser-Written Programmable Photonic Circuits}}
\TitleCitation{Toward Higher Integration Density in Femtosecond-Laser-Written Programmable Photonic Circuit}

\Author{Riccardo Albiero $^{1,2,\dagger}$, Ciro Pentangelo $^{1,2,\dagger}$, Marco Gardina~$^{1}$, Simone Atzeni~$^{2,}$*, Francesco Ceccarelli~$^{2}$ and~Roberto~Osellame~$^{2}$}

\AuthorNames{Riccardo Albiero, Ciro Pentangelo, Marco Gardina, Simone Atzeni, Francesco Ceccarelli and Roberto Osellame}
\AuthorCitation{ Albiero, R.; Pentangelo, C.; Gardina, M.; Atzeni, S.; Ceccarelli, F.; Osellame, R.}
\address{%
$^{1}$ \quad Department of Physics, Politecnico di Milano, Piazza Leonardo da Vinci 32, 20133 Milano, Italy; riccardo.albiero@polimi.it (R.A.); ciro.pentangelo@polimi.it (C.P.); marco.gardina@mail.polimi.it (M.G.)\\
$^{2}$ \quad Istituto di Fotonica e Nanotecnologie---Consiglio Nazionale delle Ricerche (IFN-CNR), Piazza Leonardo da Vinci 32, 20133 Milano, Italy; francesco.ceccarelli@cnr.it (F.C.); roberto.osellame@cnr.it (R.O.)}

\corres{Correspondence: simone.atzeni@polimi.it}

\firstnote{These authors contributed equally to this work.}

\abstract{Programmability in femtosecond-laser-written integrated circuits is commonly achieved with the implementation of thermal phase shifters.
Recent work has shown how such phase shifters display significantly reduced power dissipation and thermal crosstalk with the implementation of thermal isolation structures. However, the aforementioned phase shifter technology is based on a single gold film, which poses severe limitations on integration density and circuit complexity due to intrinsic geometrical constraints.  To increase the compactness, we propose two improvements to this technology. Firstly, we fabricated thermal phase shifters with a photolithography process based on two different metal films, namely (1) chromium for microheaters and (2) copper for contact pads and interconnections. Secondly, we developed a novel curved isolation trench design that, along with a state-of-the-art curvature radius, allows for a significant reduction in the optical length of integrated circuits. As a result, curved \ce{Cr}-\ce{Cu} phase shifters provide a compact footprint with low parasitic series resistance and no significant increase in power dissipation ($\sim$\SI{38}{\milli\watt}) and thermal crosstalk ($\sim${20}\%). These results pave the way toward the fabrication of femtosecond-laser-written photonic circuits with a steep increase in terms of layout complexity.}

\keyword{femtosecond laser micromachining; programmable photonic circuits; thermal phase shifting; universal photonic processors}

\begin{document}
\section{Introduction}

Integrated photonics is a fundamental technology for many applications, such as quantum information processing~\cite{Wang2020, Arrazola2021, Pelucchi2022} and signal routing and communication~\cite{Annoni2017, Miller2019}, where it represents a clear path toward the realization of large-scale photonic devices and networks~\cite{ Pelucchi2022, Bogaerts2020}.
Indeed, a bulk optics setup with discrete optical components shows stability and scalability limitations that become more apparent with the growth of the protocol complexity. An integrated approach, on the other hand, enables the realization of a large number of optical components on a single monolithic device, maintaining excellent interferometric stability and a small footprint. Furthermore, an appealing feature of photonic integrated circuits (PICs) is the possibility to actively reconfigure the circuit operation by relying on electrically programmable components such as phase shifters.

In recent years, there has been a growing interest toward the development of universal photonic processors (UPPs)~\cite{Ruocco2016, Harris2018, Taballione2021, Taballione2022}, which are programmable PICs that can perform any arbitrary unitary transformation on a given set of input signals. Circuit topologies for the realization of UPPs are well assessed in the literature, with two notable examples being triangular~\cite{Reck1994} and rectangular~\cite{Clements2016} meshes of reconfigurable Mach--Zehnder interferometers (MZIs). UPPs have been already demonstrated in several integrated photonics platforms such as silicon nitride~\cite{Taballione2021, Taballione2022, deGoede2022}, silica-on-silicon~\cite{Carolan2015} and femtosecond laser micromachining (FLM)~\cite{Dyakonov2018, PentangeloSPIE}. 
In particular, FLM is a versatile and cost-effective fabrication platform for implementing integrated optical devices. This technique exploits the non-linear interaction of focused femtosecond laser pulses with a transparent dielectric material to induce a permanent change in the refractive index localized in the focal volume, allowing the fabrication of waveguides by translating the sample underneath the pulsed laser. Femtosecond-laser-written waveguides in glass substrates show low optical losses (less than \SI{0.3}{\decibel\per\cm} in the visible and near-infrared~\cite{Corrielli2021}), have low birefringence (as low as \SI{1.2e-6}{}~\cite{Corrielli2018}, making them compatible with polarization-encoded protocols), and can be easily exploited for the fabrication of complex structures in a three-dimensional fashion~\cite{Crespi2016, Hoch2022}.

Reconfigurability can be achieved by leveraging the thermo-optic effect: that is, the dependence of the refractive index on temperature.
The most straightforward implementation within the FLM fabrication platform involves the integration of microheaters on the surface of the device, which act as resistive elements, allowing the dissipation of electrical power by Joule effect and a  consequent localized substrate temperature increase.
This technique is particularly effective since microheaters may be easily realized by depositing a metal film on top of the substrate, which is then patterned using the same femtosecond laser system used for the fabrication of the optical structures~\cite{Flamini2015, Ceccarelli2019}. They also offer excellent operation stability while causing no additional photon losses. On the other hand, thermal phase shifters suffer from some drawbacks, the most notable being thermal crosstalk, which is related to the heat diffusion toward waveguides different than the target one.
Waveguides can be thermally isolated by microstructuring the glass substrate and fabricating deep isolation trenches, as shown in \cite{Ceccarelli2020}, to reduce the influence of thermal crosstalk while providing orders of magnitude reduction in the dissipated power. When operated in a vacuum environment~\cite{Ceccarelli2020}, devices featuring thermal isolation structures display an even greater reduction in dissipated power while making thermal crosstalk almost negligible at the cost of a slower time response. 

Scaling up the computational power of UPPs requires an increase in both the number of optical modes and thermal phase shifters, which are practically limited to a handful and a few tens, respectively, with the FLM fabrication process adopted in~\cite{Ceccarelli2020}. Indeed, in \cite{Ceccarelli2020}, the resistive elements, the interconnections and the contact pads are all patterned on the same metal film; therefore, the aspect ratio is the only degree of freedom that can be exploited to concentrate all the electrical resistance, and thus the power dissipation, in the microheater. In essence, the metal interconnections require an aspect ratio close to 1:1 in order to reduce the parasitic series resistance, therefore occupying a large surface area on the device. This poses a severe limitation on the number of thermal phase shifters that can be manufactured on a single chip.
In addition, the overall optical losses should be kept as low as possible as the complexity of the device increases. This can be achieved by both optimizing the waveguide propagation losses and reducing the total circuit length.

To address these issues, we propose two different solutions: on the one hand, we reduce the footprint of the programmable MZI (the building block of UPPs) by both optimizing the minimum curvature radius with negligible additional losses and developing curved isolation trenches. In this way, we maintain the waveguides as  thermally isolated while fabricating MZIs with 'null' arm length, i.e., without straight waveguide segments for the phase shifters, thus saving some millimeters per cell. On the other hand, we propose the use of a two-metal planar lithography technique to fabricate thermal phase shifters. This approach allows using a metal with high electrical resistance for the fabrication of the heaters while using a low resistance metal for the realization of long and narrow metal interconnections with negligible parasitic series resistance. However, performing photolithography on a microstructured substrate forbids the use of standard liquid photoresist deposited by spin coating, which makes the development of such a technology not trivial. The process we developed makes use of a dry photoresist, which can tent over isolation structures, providing homogeneous coverage.

The manuscript is organized as follows: in \secref{sec:fabrication}, we provide a description of the fabrication process developed for the thermal phase shifters and for the curved deep isolation trenches. In \secref{sec:experimental}, we present the electrical and optical characterization and provide a performance comparison between rectangular and curved trenches with various bending radii. Lastly, in \secref{sec:discussion} we discuss the results and the prospects of this work and finally draw our conclusions in \secref{sec:conclusion}.

\section{Fabrication Improvements}\label{sec:fabrication}
\subsection{Compact Mach--Zehnder Interferometer Design}\label{subsec:compactMZI}

The following discussion will be centered on the programmable MZI, which represents the fundamental building block for universal multiport devices~\cite{Reck1994, Clements2016}.
The interferometers are fabricated in boro-aluminosilicate glass (Corning EAGLE XG, \SI{1}{\milli\meter} thick) and inscribed \SI{35}{\micro\meter} from the glass surface. Optical waveguides are fabricated using the multi-scan technique and thermal annealing~\cite{Arriola2013}. Single-mode operation has been optimized for a wavelength of \SI{940}{\nano\meter}, achieving propagation losses down to \SI{0.14}{\decibel\per\centi\metre} and negligible bending losses at curvature radii as low as $R~=~\SI{15}{\milli\meter}$. This value shows an improvement of a factor of 2 or higher with respect to the one employed in previous laser-written UPPs~\cite{PentangeloSPIE, Dyakonov2018}, and it is in line with the state of the art for the FLM platform~\cite{Lee2021}, where advanced techniques for the reduction in bending losses were introduced.

\begin{figure}[H]
    \centering
    \begin{subfigure}{.34\textwidth}
        \includegraphics[height=4cm]{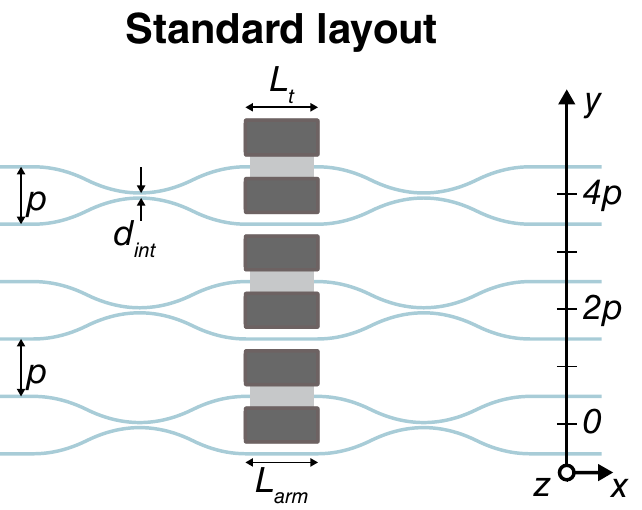}
        \caption{}
        \label{fig:design1}
    \end{subfigure}\hfill%
    \begin{subfigure}{.34\textwidth}
        \includegraphics[height=4cm]{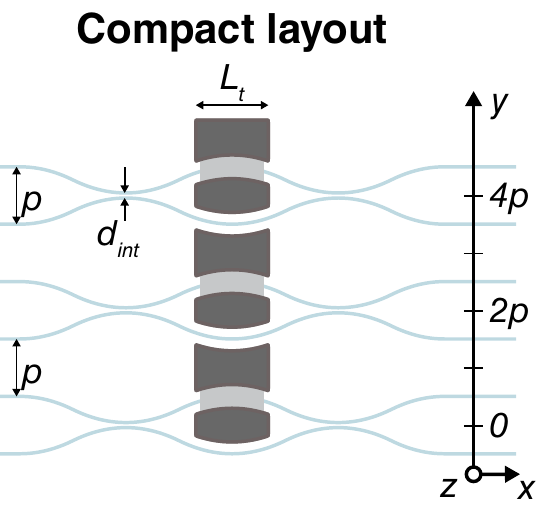}
        \caption{}
        \label{fig:design2}
    \end{subfigure}\hfill%
    \begin{subfigure}{.26\textwidth}
        \includegraphics[height=4cm]{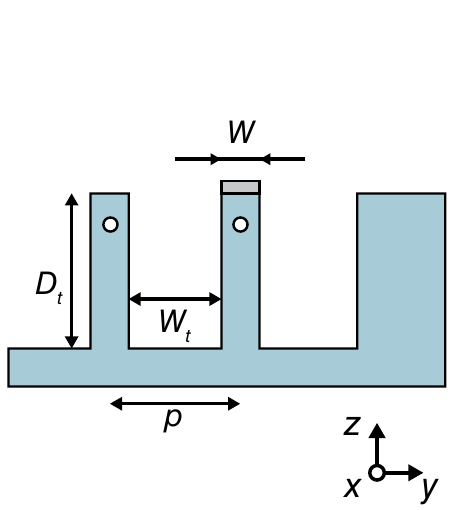}
        \caption{}
        \label{fig:crossSection}
    \end{subfigure}
    \caption{Scheme of the devices. 
    (\textbf{a}) MZIs featuring rectangular trenches and thus finite arm length. (\textbf{b}) MZIs featuring curved trenches and thus 'null' arm length. In this case, the couplers are closer and the overall circuit length is reduced. (\textbf{c}) Section view of isolation trenches. White circles are waveguide cross-sections, while the gray rectangle is the microheater cross-section.}
    \label{fig:circuit_scheme}
\end{figure}

We considered two different interferometer designs, as shown in \figref{fig:circuit_scheme}.
The first design (\figref{fig:circuit_scheme}a) comprises two balanced directional couplers realized with circular S-bend segments (curvature radius $R~=~\SI{15}{\milli\meter}$), which are connected by a straight segment with a length $L_{arm}~=~\SI{1.5}{\milli\meter}$. This MZI structure is similar to the one presented in \cite{Ceccarelli2020}, and it is used as a control sample throughout this study.
The second design (\figref{fig:design2}), on the other hand, uses circular segments, and the two balanced directional couplers are directly connected together without straight segments ($L_{arm}~=~\SI{0}{\milli\meter}$).
In order to study this compact MZI cell, we fabricated interferometers with different curvature radii, namely $R$ = 15~mm, 20 mm and 25 mm.
The pitch between the optical modes is $p~=~\SI{80}{\micro\meter}$, while the interaction distance and the interaction length of the directional couplers are $d_{int}~=~\SI{7}{\micro\meter}$ and  $L_{int}~=~\SI{0}{\milli\meter}$, respectively, for all of the interferometers.

\begin{figure}[H]
    \begin{subfigure}{.49\textwidth}
        \includegraphics[width=0.99\textwidth]{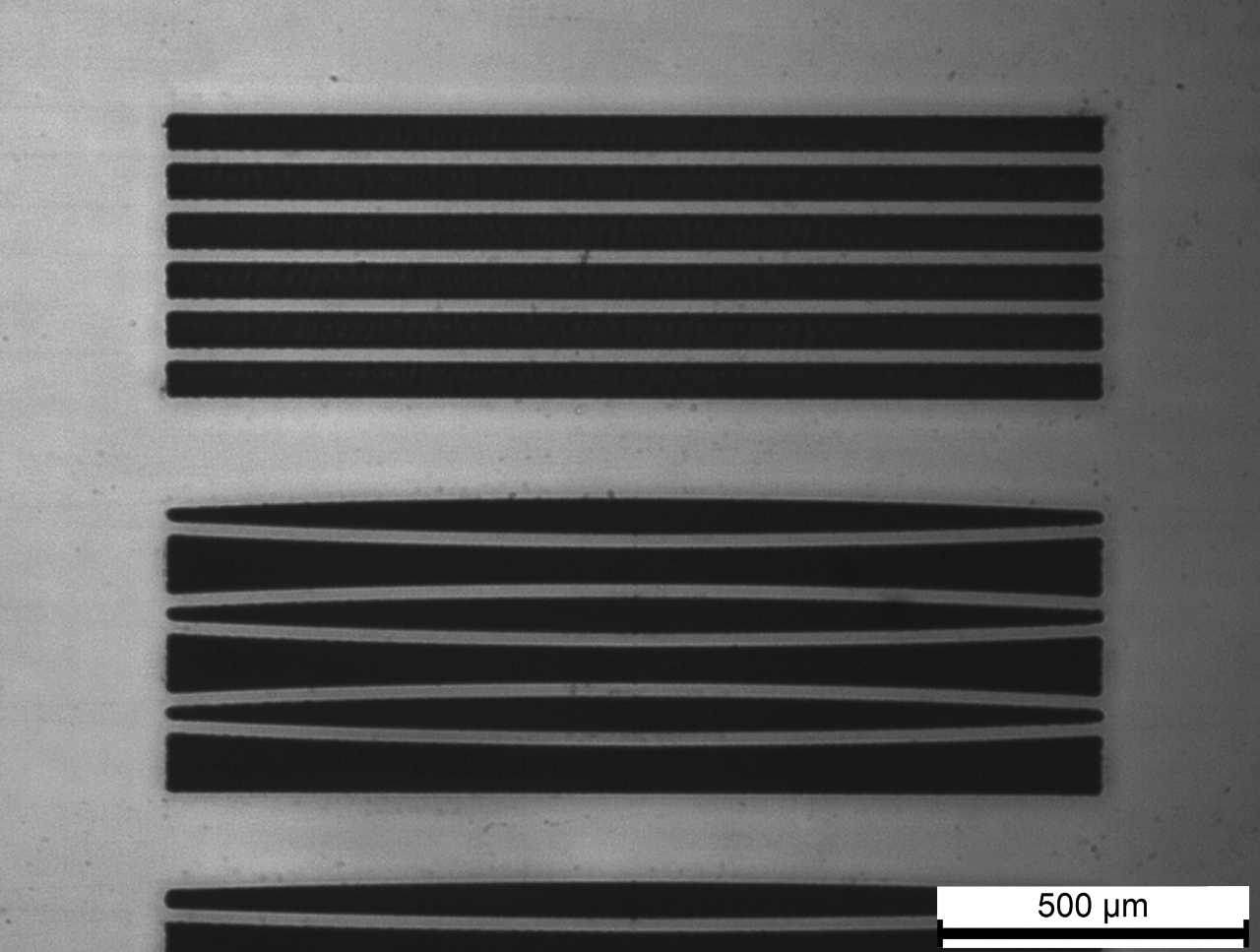}
        \caption{}
        \label{fig:litho1}
    \end{subfigure}%
    \hfill{}%
    \begin{subfigure}{.49\textwidth}
        \includegraphics[width=0.99\textwidth]{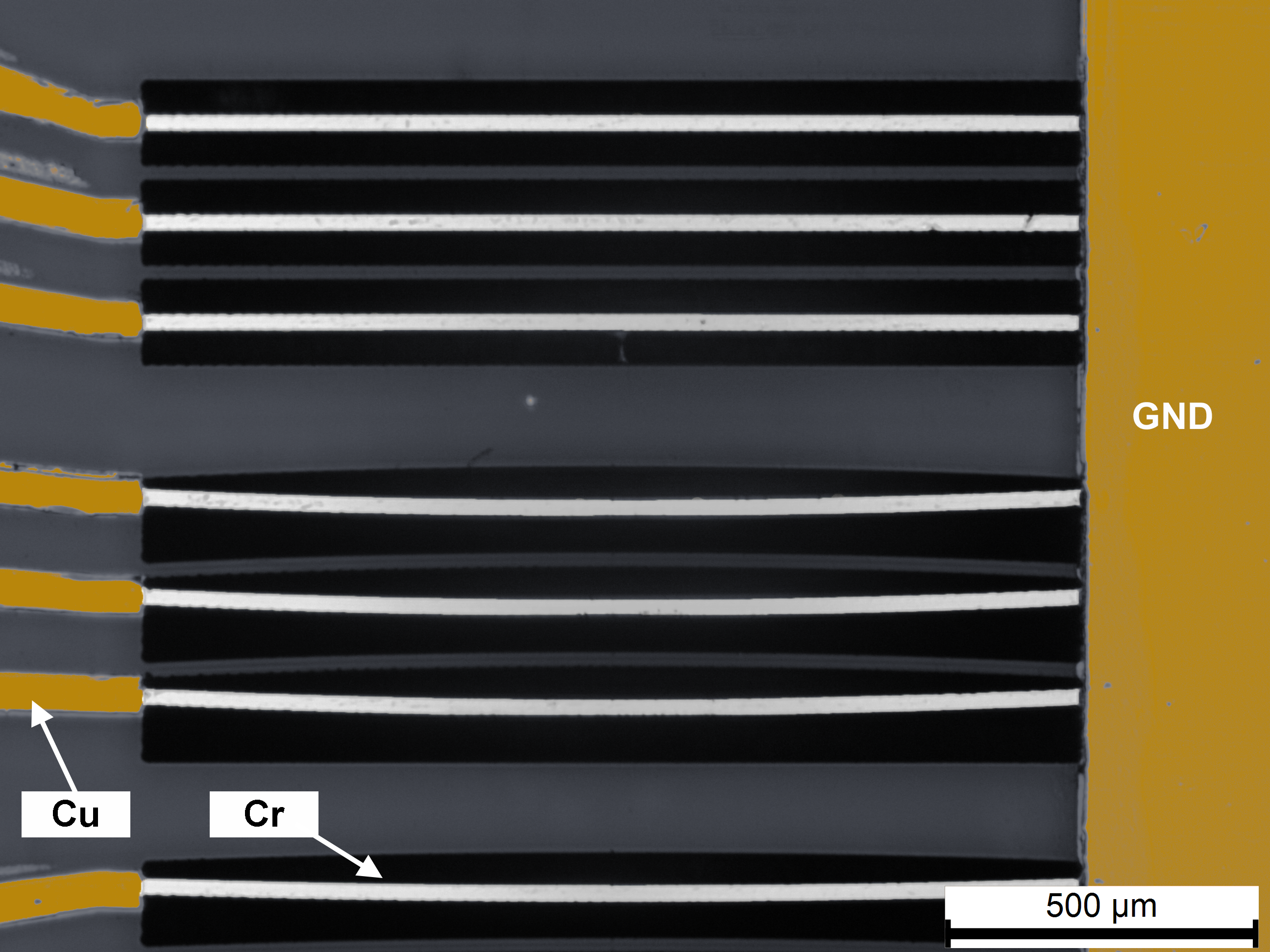}
        \caption{}
        \label{fig:litho2}
    \end{subfigure} 
    \label{fig:fabrication}
    \caption{Microscope images of microheaters before and after microheater fabrication. All images show the same MZI groups: straight arms at the top, curved ones with radius $R~=~\SI{15}{\milli\metre}$ at the bottom. (\textbf{a}) Bare substrate with isolation trenches. (\textbf{b}) Chromium microheaters and copper interconnections deposited and patterned (image in false colors).  \label{fig2}}
\end{figure}

\textls[-15]{Deep isolation trenches have been fabricated by water-assisted laser ablation~\cite{Li2013, Ceccarelli2020}. Optimized shapes have been produced for each interferometric design: standard rectangular trenches were used for the first MZI structure, while a novel curved layout was optimized for structures with null arm length. In this latter case, two different blocks are required: a biconvex and biconcave one, to fit, respectively, the inner and outer area between the two optical modes of the MZI (see \figref{fig:circuit_scheme}b). Aside from the form factor, all the trench designs have the same geometrical parameters (\figref{fig:circuit_scheme}c): a depth $D_t~=~\SI{300}{\micro\meter}$, a wall width $W~=~\SI{25}{\micro\meter}$ and a length $L_t~=~\SI{1.5}{\milli\meter}$ to accommodate thermal phase shifters of the same length. A microscope image of these structures is reported in~\figref{fig2}a.}

\subsection{Two-Metal Photolithography Technique for Thermal Phase Shifters}\label{subsec:lithography}

Chromium was chosen as the material for the resistive elements due to its high resistivity as well as its lower temperature coefficient of resistance (TCR) in the temperature range we are interested in (20 to 200 $^{\circ}$C). Similarly, copper was chosen as the material for the interconnections and the contact pads due to its significantly lower resistivity and compatibility with wire bonding.
A thickness of \SI{100}{\nano\metre} was chosen for the chromium film, considering microheater dimensions of $W_m = W = \SI{25}{\micro\metre}$ by $L_m = L_t = \SI{1.5}{\milli\metre}$.
For the interconnections, a \SI{10}{\nano\metre} thin film of titanium was used as an adhesion layer before the actual deposition of \SI{1}{\micro\metre} of copper.
These metal films were all deposited by electron-beam evaporation with an Evatec BAK 640 and patterned via photolithography and wet etching as follows.

Lithography was performed with a negative dry film photoresist (Ordyl FP 450) instead of a typical liquid photoresist deposited by spin coating, since such a technique is not ideal for rectangular and microstructured substrates due to the formation of comets, poor edge coverage and severe inhomogeneities. The \SI{50}{\micro\metre} thick dry film is laminated on the substrate by a dry film laminator at a temperature of \SI{80}{\degreeCelsius}. The thickness of the dry film allows it to 'tent' over the trenches, evenly covering the surface around the isolation structures.
The dry film is then exposed by a Heidelberg MLA100 maskless aligner with a dose of \SI{500}{\milli\joule\per\centi\metre\squared} and developed for \ang{;2;40} in a {1}\% solution of \ce{Na2CO3} at room temperature. Before proceeding with wet etching, the sample is typically placed in an argon plasma (\SI{200}{\watt} for \ang{;2;}). After etching, the photoresist mask is stripped away in a {3}\% solution of \ce{KOH}.

After deposition of the chromium film, a dry film photoresist mask covering both the microheater and interconnections is prepared. Chromium is then etched with the Microchemicals TechniEtch Cr01 etchant mixture for \ang{;;40}. Subsequently, the titanium and copper films are evaporated in the same step, and a second photoresist mask is prepared, covering only the interconnections and not the microheaters. Selective copper etching is performed by dipping the sample in a 40:1:1 ratio solution of \ce{H2O}, \ce{HCl} and {30}\% \ce{H2O2} for \ang{;2;}. The titanium thin film is then etched using the same photoresist mask with buffered oxide etch (BOE) for \ang{;;15}. The device with fully patterned microheaters and interconnections is then placed in a vacuum annealer at \SI{350}{\degreeCelsius} for \SI{1}{\hour} in order to stabilize the metal films~\cite{Ceccarelli2019} (\figref{fig2}b).

Finally, to avoid resistance drifting during operation due to the oxidation of both metals, we deposit a passivation layer of \SI{200}{\nano\metre} of \ce{SiO2} via plasma-enhanced chemical vapor deposition (PECVD) with a STS Multiplex. The silica above the copper contact pads at the sides of the sample is then opened with a final lithography step and wet etching in BOE for \ang{;2;}.

\section{Experimental Results}\label{sec:experimental}

\subsection{Electrical Measurements}\label{subsec:electricalMeas}

A total of 12 microheaters were fabricated and divided in four groups of three MZIs, as reported in~\figref{fig:circuit_scheme}. The first group corresponds to the control sample with straight thermal shifters (\figref{fig:circuit_scheme}a), while the other three groups correspond to the curved microheaters with different radii (\figref{fig:circuit_scheme}b).
The microheaters featured an average resistance of \SI{850}{\ohm} before annealing, which dropped to \SI{720}{\ohm} after annealing. These results are in perfect agreement with the sheet resistance measurements that we have performed on plain chromium films and that resulted in $R_{sh}~=~\SI{12}{\ohm\per\sq}$. On the other hand, the copper film features a sheet resistance $R_{sh}~=~\SI{0.02}{\ohm\per\sq}$, which leads to a parasitic series resistance <$\SI{5}{\ohm}$ for the adopted design and, thus, negligible with respect to the microheater resistance.

\begin{figure}[H]
    \begin{subfigure}{.49\textwidth}
        \includegraphics[width=0.99\textwidth]{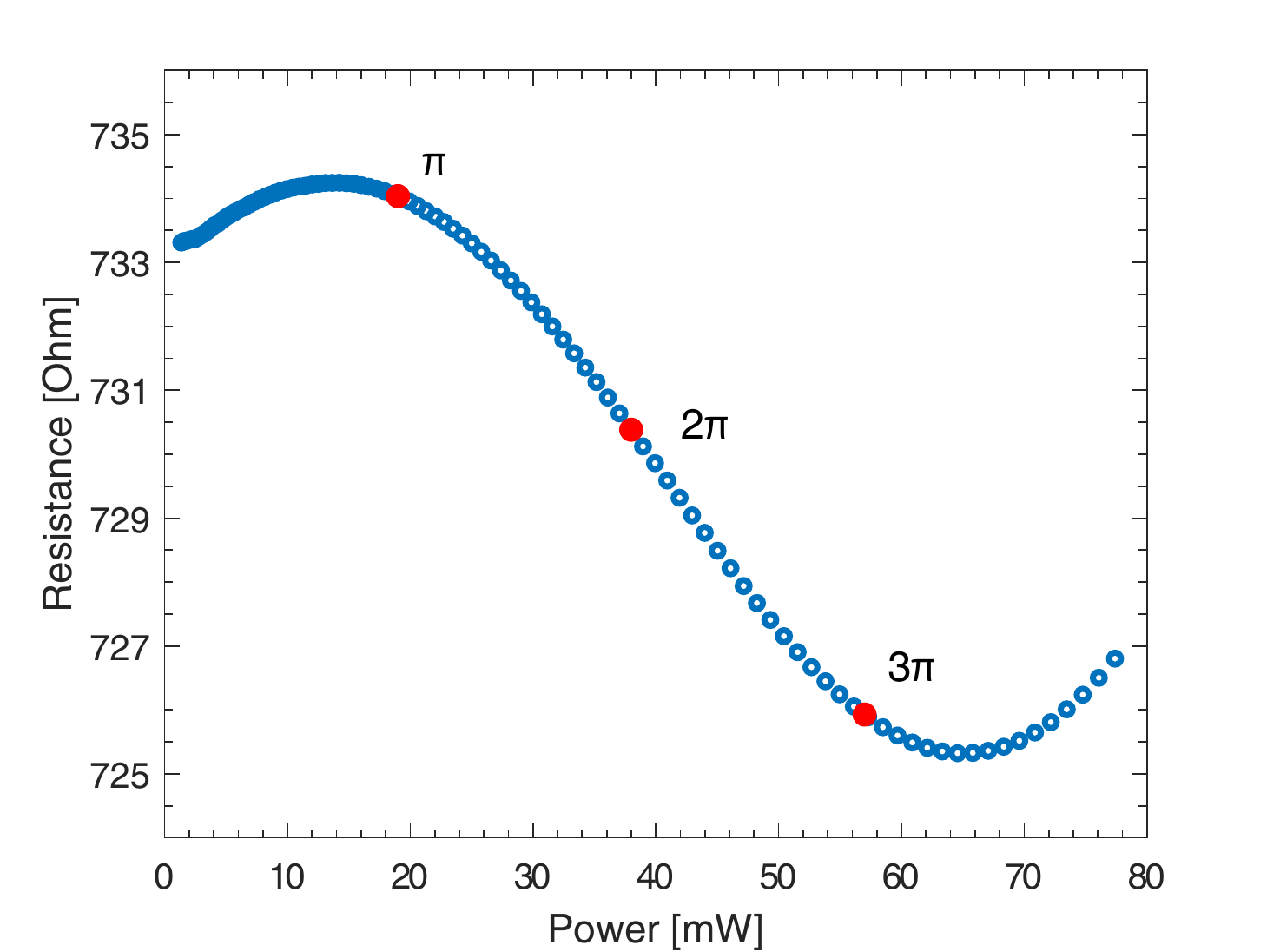}
        \caption{}
        \label{fig:rvsp}
    \end{subfigure}\hfill{}%
    \begin{subfigure}{.49\textwidth}
        \includegraphics[width=0.99\textwidth]{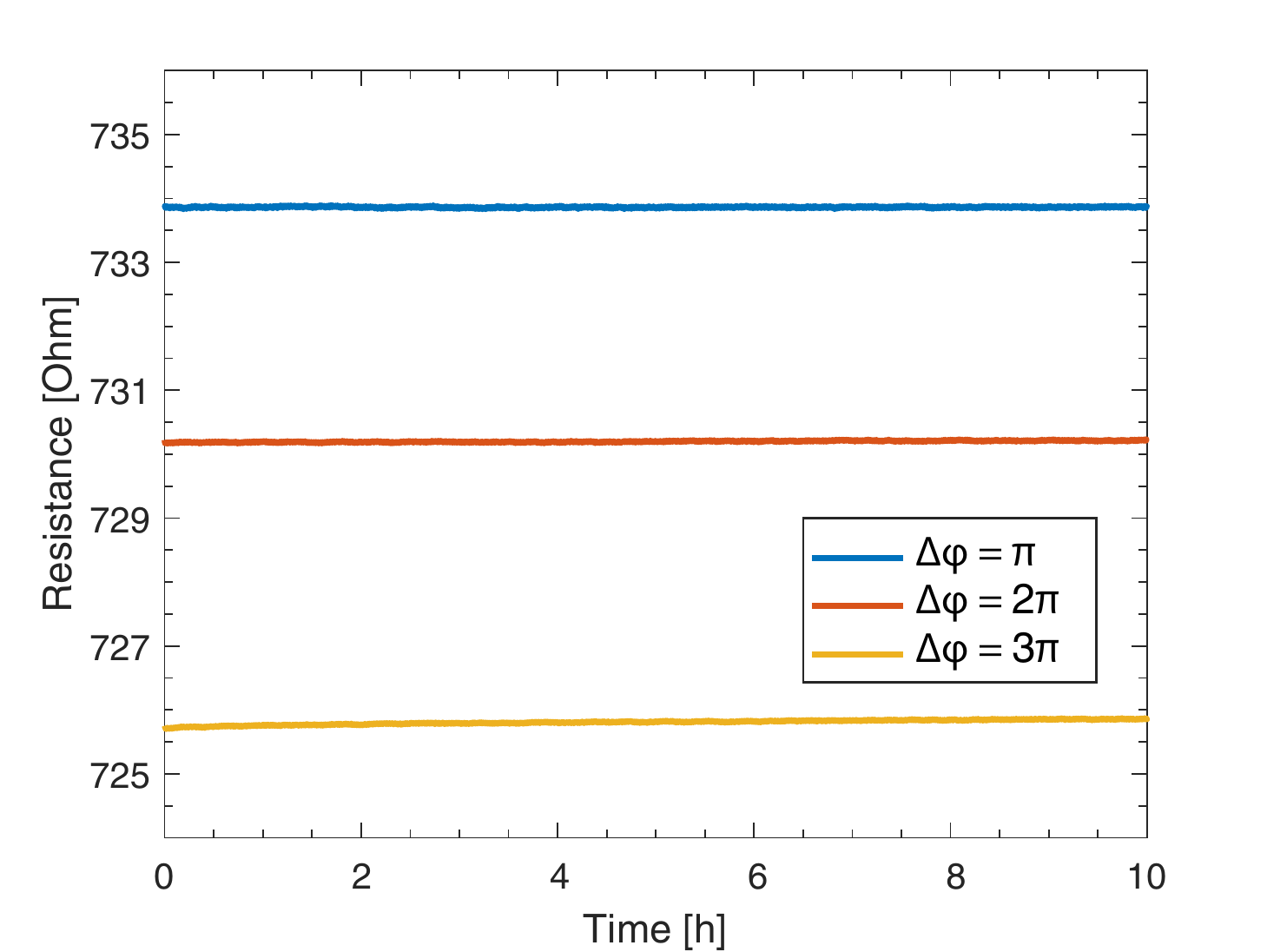}
        \caption{}
        \label{fig:stability}
    \end{subfigure}
    \caption{(\textbf{a}) Electrical characterization of the resistance versus electrical power on a curved microheater with $R = \SI{15}{\milli\metre}$; markers indicate power values corresponding to specific phase shifts. (\textbf{b})~Stability measurements performed on the same microheater at electrical powers corresponding to $\pi$, $2\pi$, and $3\pi$ phase shifts. \label{fig3}}
\end{figure}

Resistance variation of chromium within the operation temperature range was assessed by measuring it with increasing voltage up to a maximum power dissipation corresponding to a $3\pi$ phase shift. The resulting graph for a microheater featuring curved trenches with $R = \SI{15}{\milli\meter}$ can be seen in \figref{fig3}a. The maximum variation of resistance in the working temperature range is as low as {1}\% from the value measured at room temperature. This value is very low when compared to an identical gold microheater, which would vary by more than {30}\% in the same temperature range due to gold's higher TCR~\cite{Ceccarelli2019}. 
Although the curve reported in \figref{fig3}a looks peculiar for a metal, it is worth mentioning that the electrical resistivity of chromium and its non-trivial temperature dependence are heavily affected by many factors such as film thickness, the presence of contaminants, and the grain structure of the film~\cite{Marcinkowski1961, Arajs1973, Ohashi2016, Rapp1978}. However, further investigation on the resistivity of this film is beyond the scope of this article.

Finally, we performed stability measurements to verify the presence of possible resistance drifts at various dissipated powers. Such drifts can arise from either an insufficient thermal budget provided during annealing (typically resulting in a resistance decrease over time) or microheater oxidation due to defective passivation (resulting in a resistance increase over time). Microheaters featuring thermal isolation trenches were actuated at powers corresponding to $\pi$, $2\pi$ and $3\pi$ phase shifts for \SI{10}{\hour}, resulting in a maximum resistance variation of {0.02}\%. These chromium microheaters indeed do not display significant drift over time, indicating a properly working annealing and passivation. 

\subsection{Optical Measurements}\label{subsec:opticalMeas}
The performance of all the fabricated devices was characterized in terms of power dissipation, crosstalk and dynamic response. We report data only for the last microheater of each group (see~\figref{fig:circuit_scheme}). However, no significant deviations were observed by actuating the other microheaters. All the results of the characterization are reported in~\tabref{tab:summary} and will be detailed in the following sections.

\begin{table}[H]
\caption{Summary of experimental measurements. Cell length, dissipated power, thermal crosstalk ($d~=~2p$) and response times for both large and small signal regimes are reported for all MZI groups considered in this work. }
\label{tab:summary}
 
\begin{adjustwidth}{-\extralength}{0cm}
\begin{minipage}{\fulllength}
\setlength{\cellWidtha}{\textwidth/6-2\tabcolsep+0.50in}
\setlength{\cellWidthb}{\textwidth/6-2\tabcolsep-0in}
\setlength{\cellWidthc}{\textwidth/6-2\tabcolsep-.10in}
\setlength{\cellWidthd}{\textwidth/6-2\tabcolsep-.10in}
\setlength{\cellWidthe}{\textwidth/6-2\tabcolsep-.20in}
\setlength{\cellWidthf}{\textwidth/6-2\tabcolsep-.10in}
		\begin{tabularx}{\textwidth}{>{\raggedright\arraybackslash}m{\cellWidtha}>{\centering\arraybackslash}m{\cellWidthb}>{\centering\arraybackslash}m{\cellWidthc}>{\centering\arraybackslash}m{\cellWidthd}>{\centering\arraybackslash}m{\cellWidthe}>{\centering\arraybackslash}m{\cellWidthf}}
			\toprule
        \textbf{Isolation Design} & \textbf{Cell Length (mm) }& \boldmath{$P_{2\pi}$} \textbf{(mW)} & \boldmath{$\Delta\varphi_{ct}^{2p}/\Delta\varphi$} \textbf{(\%)} & \boldmath{$\tau_{large}$} \textbf{(ms)} & \boldmath{$\tau_{small}$} \textbf{(ms) }\\
        \midrule
        Rectangular trenches,  $R_\infty$                      &  7.42 & 35.02 & 21.31 & 34.4 & 57.7 \\
        Curved trench,         $R = \SI{25}{\milli\meter}$     &  7.64 & 38.23 & 20.17 & 29.9 & 52.5 \\
        Curved trench,         $R = \SI{20}{\milli\meter}$     &  6.83 & 38.33 & 20.07 & 28.9 & 55.4 \\
        Curved trench,         $R = \SI{15}{\milli\meter}$     &  5.92 & 38.10 & 18.64 & 29.8 & 50.3 \\
        \bottomrule
    \end{tabularx}
\end{minipage}
\end{adjustwidth}
\end{table}

\subsubsection{Dissipated Power}\label{subsubsec:dissipatedPower}
For a given MZI, the phase difference $\Delta\varphi$ induced between its arms by the microheater can be expressed as~\cite{Flamini2015}:
\begin{equation}
\Delta\varphi = \varphi_0 + \alpha P
    \label{eq:dissipated_power}
\end{equation}
where $\varphi_0$ is the phase difference between the arms of the interferometer when no power is dissipated, $\alpha$ is the interferometer tuning coefficient and $P$ is the dissipated power.
In particular, $P_{2\pi}$ is the electrical power that a thermal phase shifter has to dissipate to induce a $\Delta\varphi = 2\pi$ phase shift and achieve full reconfigurability.

In \cite{Ceccarelli2020} was demonstrated that the use of deep isolation trenches provides a significant improvement in terms of $P_{2\pi}$ with respect to a planar device.
In this study, we compared the performance of the curved isolation trenches with that of the rectangular ones in order to determine how the different geometry affects the isolation performance.
$P_{2\pi}$ was characterized experimentally for all of the MZIs with the completely curved design and the control sample in order to be able to draw a comparison with the results in the literature. 

We measured a power dissipation $P_{2\pi}= \SI{38}{\milli\watt}$, which is constant among all the MZIs fabricated with the curved design, indicating that $P_{2\pi}$ does not depend on the curvature radius. This result is not trivial. Indeed, while the width $W_t$ of the waist of each block is identical to that of the rectangular trenches, the minimum width of the curved trenches varies by a factor of two (from a minimum of \SI{13}{\micro\metre} to a maximum of \SI{27}{\micro\metre}) as the radius increases from $R$~=~15 to 25~mm, potentially impacting on the isolation capability of the structure. Furthermore, the $P_{2\pi}$ values measured for the control sample, namely \SI{35}{\milli\watt}, are comparable to the results obtained with curved trenches. This result indicates that $P_{2\pi}$ is not significantly affected by trench shape, allowing the realization of compact MZIs with nearly the same dissipated power as the previous schemes.

\subsubsection{Thermal Crosstalk}

Under the assumption of linear dependence of the refractive index on the temperature change of the substrate, the phase shift induced by thermal crosstalk on a given MZI can be modeled as~\cite{Ceccarelli2020}: 
\begin{equation}
   \Delta\varphi_{ct} = \sum_i{\alpha_i P_i},
\end{equation}
where the sum is performed on all neighboring MZIs, $P_i$ is the power dissipated on each microheater and $\alpha_i$ represents the crosstalk tuning coefficients between each adjacent phase shifter and the target MZI.
We can evaluate the effect of thermal crosstalk by measuring the phase $\Delta\varphi_{ct}$ induced on a target MZI while dissipating power on its neighboring microheaters at various distances. The magnitude of thermal crosstalk can then be evaluated in terms of the ratio of phase induced by a neighboring heater and by the target heater when actuated with the same power $P = P_i$.
As expected, the additional thermal isolation between waveguides in substrates featuring deep trenches helps to dampen additional phase contributions from nearby phase shifters, reducing thermal crosstalk effects when compared to planar substrates.
Thermal crosstalk was measured for all four MZI designs for the first and second neighbors (distances of $2p$ and $4p$; see~\figref{fig:circuit_scheme}a,b).

\begin{figure}[H]
    \begin{subfigure}{.49\textwidth}
        \includegraphics[width=0.99\textwidth]{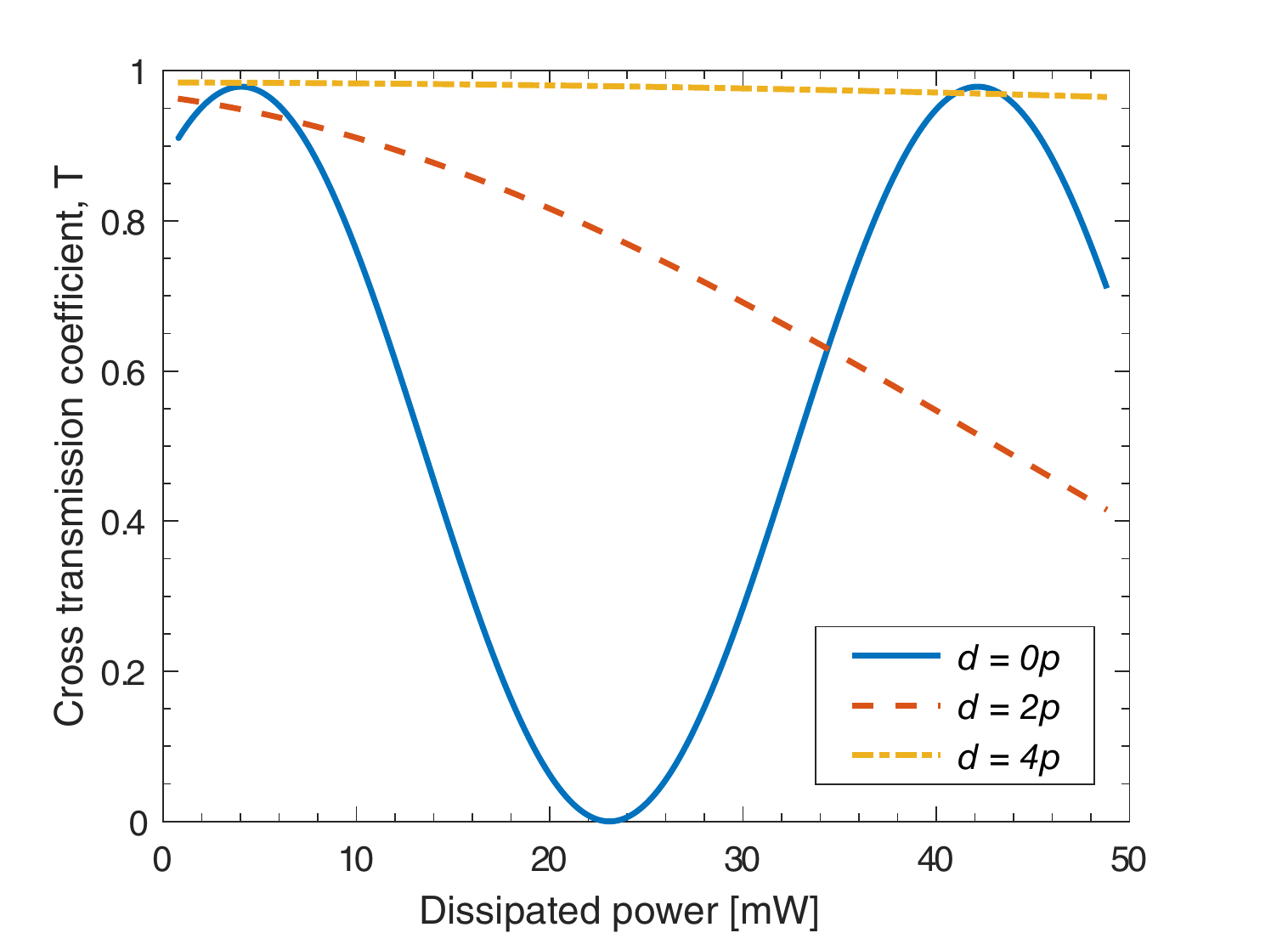}
        \caption{}
        \label{fig:crosstalk}
    \end{subfigure}\hfill{}%
    \begin{subfigure}{.49\textwidth}
        \includegraphics[width=0.99\textwidth]{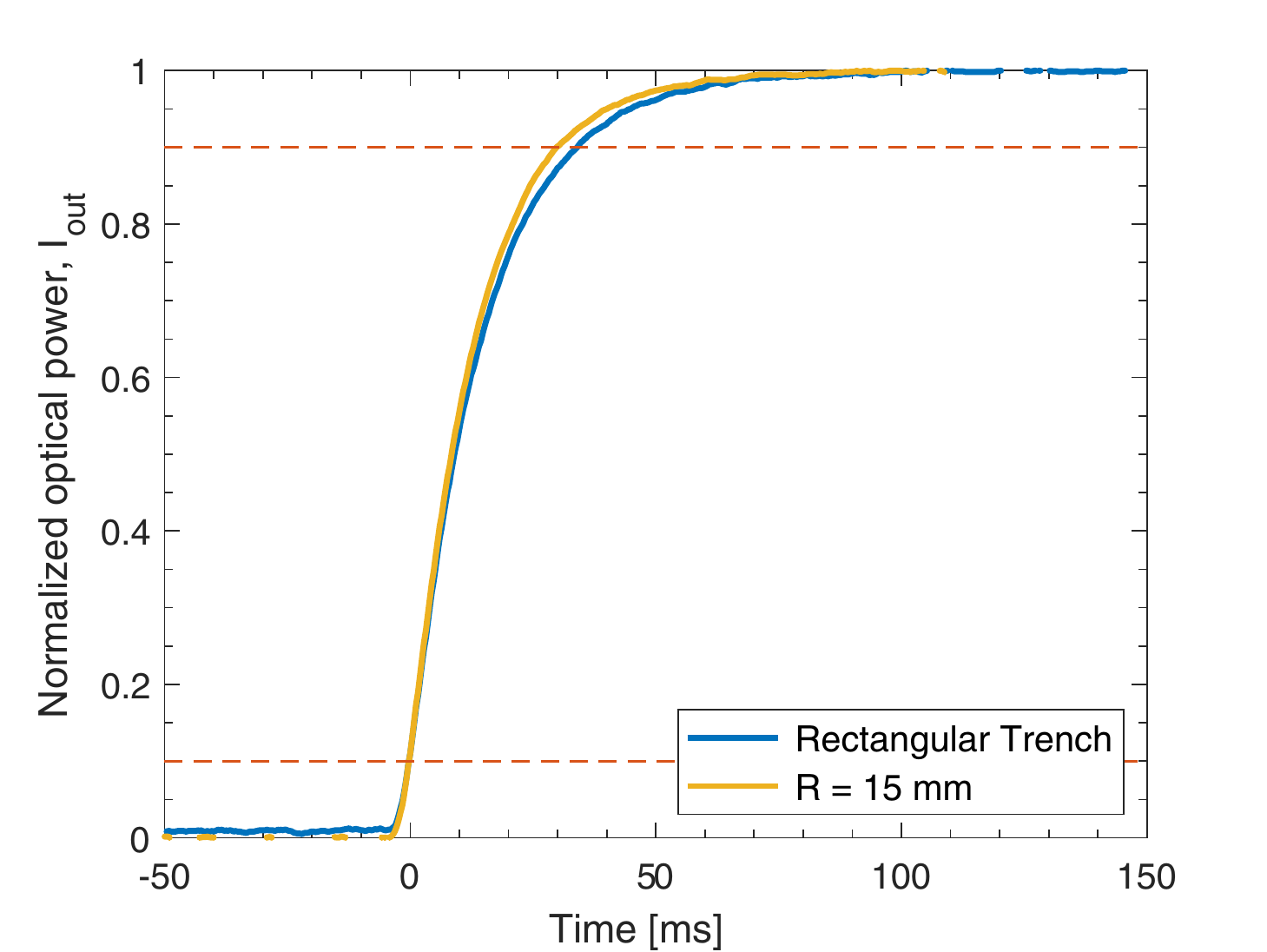}
        \caption{}
        \label{fig:rising_time}
    \end{subfigure}
    \caption{(\textbf{a}) Normalized optical power as a function of dissipated power on target MZI and neighboring MZIs measured on devices with curved trenches of radius $R~=~\SI{15}{\milli\metre}$. (\textbf{b}) Normalized optical power as a function of time for the rectangular trench ($L_{t} = \SI{1.5}{\milli\meter}$) and curved trench ($R~=~\SI{15}{\milli\meter}$). \label{fig4}}
\end{figure}

We report in~\figref{fig4}a the thermal crosstalk measurements for the curved phase shifter with radius $R~=~\SI{15}{\milli\metre}$. We  observe a similar isolation performance for all microstructures, with $\sim$20\% phase induced on the first neighbor at a distance $2p = \SI{160}{\micro\metre}$, suggesting an independence of crosstalk against both curvature radius and geometry, similarly to power dissipation measurements. This result was again not trivial, especially considering that in the biconvex design, trench widths are as low as \SI{13}{\micro\meter} in the tails.

\subsubsection{Dynamic Response}
Dynamic response is another key performance metric for programmable circuits. While thermal isolation in general reduces both the dissipated power $P_{2\pi}$ and the thermal crosstalk, it has been shown to have a detrimental effect on the time response of the device~\cite{Pentangelo2021, Ceccarelli2020}.
Rise and fall times are computed as the interval in which the signal changes from 10\% to 90\% and vice versa of the steady-state optical power and were assessed for both the large and small signal regime. In the former case, we measured the switching time for a complete optical switch by applying a phase shift of $\Delta\varphi = \pi$. In the latter one, we measured the switching time for small phase variations $\delta\varphi$ causing a variation of the normalized optical power around the interferometer balanced working point of about {5}\%,~namely:
\begin{equation}
    \label{eq:small_signal}
    \Delta\varphi_{small} = \pi/2\pm\delta\varphi
\end{equation}
where $\pi/2$ is the phase shift to bias the MZI in a balanced state with a transmission coefficient $T = 0.5$.

The leading edges for the large signal regime are shown in \figref{fig4}b.
In this section, we will present the measurements for the heating transients, but it is worth noting that both the heating (rising) and cooling (falling) processes have comparable switching times as consistent with the theory~\cite{Pentangelo2021}. 
Similarly to the considerations we made in the previous sections, the rising times we measured are equivalent for all the curvature radii to a value of 29.5 $\pm$ 0.4 ms. Moreover, these values are comparable to the switching times measured for the rectangular trenches, namely~\SI{34.4} {\milli\second}. 
A similar behavior was observed also for the small signal response, and the values are reported in~\tabref{tab:summary}.
As a final remark, we note that at a closer observation, all previous experimental results indicate coherently a slightly reduced isolation between the two arms of a single MZI with the curved trenches (a slightly higher $P_{2\pi}$ and slightly lower response times). However, these small variations are negligible if compared to the great advantage that the curved design provides in terms of device losses. In fact, given that the propagation losses (expressed in \SI{}{\decibel}) scale linearly with the device length, the curved design provides a significant {20}\% reduction.

\section{Discussion}\label{sec:discussion}
The use of a two-metal photolithographic process for the fabrication of thermal shifters allows us to overcome the geometrical constraints imposed by a single-material approach~\cite{Ceccarelli2020}. As a result, the microheaters are made of a metal with high electrical resistivity, namely chromium, while the interconnections are made of a metal with low resistivity, namely copper.
We successfully fabricated \ce{Cr}-\ce{Cu} thermal phase shifters and demonstrated that after several hours of operation, the shifters maintain high stability in time with no drawbacks with respect to the previous technology. Such a feature has paramount importance when the device is employed in quantum photonics experiments, whose duration can be even in the range of tens of hours. In addition, the chromium microheaters feature a low TCR, which is important to develop simple yet accurate calibration protocols for complex circuits.

On the other hand, we optimized a novel deep isolation curved trench design, allowing us to  thermally isolate waveguides in MZIs with zero arm length. This design, along with a state-of-the-art curvature radius, allowed us to demonstrate a MZI cell featuring a total length of \SI{5.92}{\milli\metre} (see~\tabref{tab:summary}). These results represent the state of the art for programmable PICs fabricated through FLM with a total loss per unit cell as low as \SI{0.08}{\decibel}.

Finally, we showed how these two approaches can be combined to create compact and fully reconfigurable MZIs. We tested the curved design for different curvature radii (from $R~=~\SI{15}{\milli\metre}$ to $R~=~\SI{25}{\milli\metre}$) and found independence from curvature radius in terms of dissipated power, thermal crosstalk and dynamic response. Moreover, the new curved design shows performances that are equivalent to the one presented in~\cite{Ceccarelli2020} (see \tabref{tab:summary}).

This approach represents a starting point for further compactifying the UPP building-block design in which a thermal shifter (usually referred to as external phase shifter) is typically fabricated also at the input or output of the MZI~\cite{Clements2016}.
Actually, Walmsley et al. have already proposed a scheme~\cite{Walmsley2021} in which the external phase shifter is moved on top of the second arm of the interferometers, resulting in two thermal phase shifters integrated on top of the interferometer arms. In this case, an experimental realization of this layout can be made with minor changes to the procedure presented in this manuscript, specifically by changing the photoresist masks for the fabrication of thermal phase shifters while the waveguide fabrication process and substrate microstructuring process remain unchanged.
When scaling the complexity of the circuit, i.e., increasing the number of modes of the $N\times N$ UPP mesh, a significant advantage is gained in terms of insertion losses of the whole device. By combining the design proposed in~\cite{Walmsley2021} and our curved phase shifter approach, it would lead to \SI{3}{\milli\metre\per\cell} saving, resulting in a reduction in the propagation distance of some centimeters even with a few cascading MZIs. We also want to emphasize that this analysis is not limited to UPPs but can be applied to application-specific PICs with any~layout.

\section{Conclusions}\label{sec:conclusion}
In this work, we presented two solutions for further scaling the integration density and complexity of programmable FLM integrated optical circuits: a two-metal planar lithography technique to fabricate thermal phase shifters and the implementation of curved deep isolation trenches.
On the one hand, the new photolithographic process allows for an increased integration density of phase shifters fabricated on-chip. Moreover, using two metals allowed for a reduction in the parasitic series resistance of the contact pads while increasing the overall thermal and temporal stability of the microheaters and maintaining low $P_{2\pi}\sim\SI{38}{\milli\watt}$ and thermal crosstalk of about $\sim$20\%.
On the other hand, the implementation of curved deep isolation trenches allows for up to a {20}\% reduction in circuit length and thus a similar reduction in propagation losses. In particular, when compared to standard rectangular trenches, they take the same amount of fabrication time while enabling significantly more compact circuit layouts.
These results pave the way for increasingly more compact, stable and low-loss laser-written UPPs and PICs in general without significant compromises in terms of performance and fabrication time.

\vspace{6pt} 

\authorcontributions{Conceptualization, S.A., F.C., R.O.; optical circuit fabrication, R.A., S.A; electrical circuit fabrication, C.P., M.G., F.C.; experimental characterization, R.A., C.P., M.G.; project administration and funding acquisition, R.O.; original draft preparation, R.A., C.P. All the authors contributed to the review and editing of the manuscript.}

\funding{This research was funded by the European Union’s Horizon 2020 research and innovation program through the Future and Emerging Technologies (FET) project PHOQUSING (Grant Agreement No. 899544) and through the European Research Council (ERC) project CAPABLE (Grant Agreement No. 742745). R.A. acknowledges funding of his PhD fellowship by Thales Alenia Space Italia s.p.a.}
\acknowledgments{This work was partially performed at PoliFAB, the
micro- and nanofabrication facility of Politecnico di Milano (\href{https://www.polifab.polimi.it/}{\url{www.polifab.polimi.it}}) (visited on 18/07/2022). The authors would like to thank the PoliFAB staff for the valuable technical support. The authors would like also to thank Emanuele~Urbinati for his preliminary work on dry resist photolithography.}

\conflictsofinterest{The authors declare no conflict of interest.} 

\begin{adjustwidth}{-\extralength}{0cm}

\reftitle{References}




\end{adjustwidth}
\end{document}